One Year in the Life of Young Suns: Data Constrained Corona-Wind Model of $\kappa^1$ Ceti

*Vladimir S. Airapetian[1,2], Meng Jin[3], Theresa Lueftinger[4], Sudesha Boro Saikia[5], Oleg Kochukhov[6], Manuel Guedel[5], Bart Van Der Holst[7], W. Manchester IV[7]*

[1]American University, DC, [2]NASA GSFC/SEEC, [3]SETI, [4]ESA, [5]University of Vienna, Austria, [6]Uppsala University, Sweden, [7]University of Michigan

Abstract

The young magnetically active solar-like stars are efficient generators of ionizing radiation in the form of X-ray and Extreme UV (EUV) flux, stellar wind and eruptive events. These outputs are the critical factors affecting atmospheric escape and chemistry of (exo)planets around active stars. While X-ray fluxes and surface magnetic fields can be derived from observations, the EUV emission and wind mass fluxes, Coronal Mass Ejections and associated Stellar Energetic Particle events cannot be directly observed. Here, we present the results of a three-dimensional magnetohydrodynamic (MHD) model with inputs constrained by spectropolarimetric data, HST/STIS Far UV, X-ray data data and stellar magnetic maps reconstructed at two epochs separated by 11 months. The simulations show that over the course of the year, the global stellar corona had undergone a drastic transition from a simple dipole-like to a tilted dipole with multipole field components, and thus, provided favorable conditions for Corotating Interaction Events (CIRs) that drive strong shocks. The dynamic pressure exerted by CIRs are 1300 times larger than those observed from the Sun and can contribute to the atmospheric erosion of early Venus, Earth, Mars and young Earth-like exoplanets. Our data-constrained MHD model provides the framework to model coronal environments of G-M planet hosting dwarfs. The model outputs can serve as a realistic input for exoplanetary atmospheric models to evaluate the impact of stellar coronal emission, stellar winds and CIRs on their atmospheric escape and chemistry that can be tested in the upcoming JWST and ground-based observations.

1. Introduction.

The "Sun-In-Time" and recent multi-wavelength observations performed with Kepler, TESS, XMM-Newton and ground-based observations suggest that young (the first 0.5 Gyr) solar analogs (G-type main sequence stars) are magnetically active stars (Gudel 2007; 2020; Airapetian et al. 2020). Their magnetic activity is manifested in the form of large starspots covering up to 10% of a stellar surface, strong surface magnetic fields up to a few hundred Gauss, dense and hot X-ray bright corona, massive fast winds and frequent flare activity (Sanz-Forcada et al. 2011; 2019; Gudel 2007; Maehara et al. 2012; Kochukhov et al. 2020; Notsu et al. 2019; Airapetian et al. 2020). The characterization of coronal activity and its output, stellar winds will help in understanding the evolution of angular momentum of stars with age as fast and massive stellar winds carry away significant angular momentum of cool stars, and thus contribute to magnetic braking, which in turn controls stellar magnetic dynamo and its activity (Gallet et al. 2017; O'Fionnagain & Vidotto 2018; Brun et al. 2017).

Recent studies recognized the critical impact of stellar magnetic activity on atmospheric evolution of exoplanetary environments as a factor of their habitability (see Cohen et al. 2014; 2018; Cunz and Guinan 2016; Dong et al. 2018; Yamashiki et al. 2019; Airapetian et al. 2016; 2017; 2020 in references herein). The accurate knowledge of stellar XUV [X-ray (0.1-10 nm) and Extreme UV fluxes (EUV, 10-91.2 nm)] and stellar wind fluxes are critical inputs for the atmospheric escape and chemistry models (Cohen et al. 2014; Cuntz and Guinan 2016; Johnstone et al. 2018; 2019; Airapetian et al. 2017; 2020; Garcia-Sage et al. 2017; Vidotto & Cleary 2020). XUV radiation deposits heating via photoinization of exoplanetary atmospheres that increases neutral temperature with overall expansion transitioning at some point to hydrodynamic escape from the planetary atmospheres (Jonstone et al. 2019). These ionizing fluxes can also erode planetary atmospheres via production of photoelectrons that drive polarization electric field, the source of ion outflow. Both effects can be instrumental in driving massive winds from terrestrial type exoplanets, and therefore negatively impact exoplanetary habitability. Massive and magnetized stellar winds from young stars can exert dynamic pressure on exoplanetary magnetospheres, and thus leading to the expansion of the the polar cap area as well as generate ionospheric currents that can dissipate their energy via Joule heating, which is another factor of atmospheric escape via thermal plasma expansion (Cohen et. al. 2014; 2018). Thus, accurate knowledge of the EUV and wind fluxes are critical factors in assessing the atmospheric escape rates, and thus atmospheric evolution of Venus, Earth and Mars and exoplanets. Also, stellar winds carry away significant angular momentum of young cool stars, and thus control evolution of stellar rotation rate, which in turn governs stellar magnetic activity (Weber & Davis 1967; Sakurai 1985, Kawaler 1988).

While the X-ray portion and short-wavelength portion of EUV flux (up to 30 nm) can be derived from direct soft X-ray observations, emission measure distribution and in a few cases of very close-by stars direct EUV observations (for example, EUVE mission), the total EUV flux longer than 30 nm is heavily absorbed by neutral hydrogen of the interstellar medium (ISM) and thus cannot be reliably determined (Ribas et al. 2005; Sanz-Forcada et al. 2011; Duvvuri et al. 2021).

Also, tenuous stellar winds cannot be directly observed and characterized, but can be empirically constrained using astrospheric Ly-alpha measurements, radio observations and other proxies (Wood et al. 2005; Johnstone et al. 2015; Fichtinger et al. 2017; Jardine and Collier-Cameron 2019). Wood et al. (2005) found that the mass loss rates of winds can be scaled with the X-ray coronal flux of many magnetically active solar-like stars, which suggests that the stellar coronal heating and stellar wind acceleration are physically connected phenomena (Cranmer et al. 2015; Airapetian and Cuntz 2015).

On the Sun, the origin of supersonic and magnetized wind is physically linked to the presence of hot, 1-5-2 MK solar coronal plasma. If the slow component (450 km/s) of the solar wind can be understood as driven by thermal pressure gradient of the hot 1.5-2 MK solar corona by a simple steady-state Parker's model (1958), the acceleration of the fast wind (750 km/s) emanated from coronal holes was left unspecified. This question was later discussed in a number of studies that suggested an additional momentum term via Alfvén waves to address this problem (Usmanov et al. 2000; Ofman 2010; Cranmer et al. 2015; Airapetian and Cuntz 2015). Thus, theoretical modeling of stellar coronae that predicts X-ray coronal flux can be used as one of the methods to characterize the properties of stellar winds.

While it is generally understood that the magnetic field is the ultimate source of energy for coronal heating and wind acceleration, the details of how magnetic energy is transferred from the solar photosphere into the corona remain unsolved. Two major mechanisms have been invoked to explain the coronal heating as traced by EUV lines. The first mechanism assumes that quasi-static small-scale (granular) photospheric motions tangle, twist, and braid the magnetic field, building up magnetic free energy, which is then converted into heat via miriads of magnetic reconnection events referred to as nanoflares (e.g., Klimchuk 2006; Parker 1972 , Parker 1988, Dahlburg et al. 2016). The second heating mechanisms involves the generation and propagation of magnetohydrodynamic (MHD) waves, the Alfvén or magnetosonic waves driven by photospheric motions depositing their energy into the chromosphere, transition region and corona via resonant absorption, phase mixing, or turbulent dissipation (Osterbrock 1961; Mathiodakis et al. 2013; Cranmer and Winebarger 2019; Banerjee et al. 2009, Hahn & Savin 2013). Recent solar observations provide evidence for energy transport via upward propagating large amplitude Alfvén waves in the solar chromosphere, transition region and corona (Tomczyk et al. 2007; De Pontieu et al. 2007; 2015; Jess et al. 2009; McIntosh et al. 2011; Mathiodakis et al. 2013). The wave amplitudes are inferred from the Doppler line-of-sight velocity perturbation and non-thermal broadening of optically thin emission lines. Recently, Grant et al. (2018) presented evidence of Alfvén wave dissipation at small scales as expected from the MHD wave theory using the results of high spatial resolution observations of the chromospheric layers associated with the umbral boundary of a sunspot.

Large non-thermal broadening of emission lines forming in the upper atmosphere has also been detected in a number of active cool main-sequence stars as well as giant and supergiant stars [Wood et al. 1997; Robinson et al. (1998); Peter (2006); review in Airapetian and Cuntz (2015)]. In most cases, the non-thermally broadened emission lines show enhanced wings so that the line profile can be fit by two Gaussian profiles. While these features may also be interpreted as a signature of microflaring activity as suggested by Wood et al. (1997), they have been observed in late-type (K and M) giant stars that show no signatures of eruptive activity. Carpenter and Robinson (1997); Airapetian et al. (2000) suggested that broad and enhanced wings of optically thin emission lines can be represented by a large-scale supersonic turbulent motions, which are anisotropically distributed either along or perpendicular to the LOS. Supersonic turbulent motions can be attributed to unresolved motions associated with upward and/or downward propagating large-amplitude (non-linear) Alfvén waves that can efficiently dissipate in the solar and stellar upper atmosphere via shocks (Suzuki and Inutsuka 2006; de Pontieu et al. 2015; Airapetian et al. 2000; 2010; 2015; Shoda et al. 2021; Magyar et al. 2021). These signatures can also be formed via interaction of upward with downward propagating Alfvén waves forming due to reflection from the atmospheric regions where the gradient of the Alfvén velocity is comparable or exceeds the Alfvén wave frequency or in coronal loops (Verdini & Velli 2008; Cranmer and Winebarger 2020 and references herein). Moreover, recent Parker Solar Probe observations of the solar wind at $36 R_\odot$ suggest highly dynamic environments driven by a spectrum of Alfvénic fluctuations and "are not indicative of impulsive processes in the chromosphere or corona" (Squire et al. 2020; Chen et al. 2020).

To model X-ray emission of the corona of the Sun and solar-type stars, reseachers applied two classes of three-dimensional (3D) magnetohydrodynamic (MHD) numerical models. The first class of models assumes ad hoc thermally driven nearly isothermal corona and solar and stellar winds (van der Holst et al. 2007; Cohen et al. 2007; Vidotto et al. 2012; 2015; Do Nascimento et al. 2016; Alvarado-Gómez et al. 2016a,b; Lynch et al. 2016; 2019). However, these polytropic models cannot quantitatively address data from the solar EUV and X-ray emission, and thus require a more accurate equation for the energy transport. The second class of MHD coronal and wind models solves the full energy equation (so called thermodynamic MHD) that specifies the heating and cooling terms of the solar atmosphere and the wind. Here, the heating terms are specified either as *ad hoc* heating term that varies with height or due to the dissipation of turbulent cascade introduced by Alfvén waves while cooling terms include thermal conduction and radiative cooling via optically thin emission lines (Lionello et al. 2009; van der Holst 2014; Airapetian and Usmanov 2016; Oran et al. 2017; Usmanov et al. 2000; 2019; Alvarado-Gómez et al. 2016a,b; Boro Saikia et al. 2020; Reville et al. 2020). An advantage of a 3D MHD code, *AWSoM*, is in modeling the solar coronal X-ray and EUV fluxes (Power et al. 1999) by extending the lower boundary from the inner corona to the upper chromosphere, where the Poynting flux of Alfvén waves can be constrained directly from the spectroscopic observations. This model is thus can be constrained by the data supplied from the solar chromospheric emission lines (Oran et al. 2013; 2017). The model assumes that the solar corona is heated predominantly by the dissipation of Alfvén waves and can successfully reproduce the overall global structure of the solar corona, the X-ray and EUV emission from active regions and the solar wind temperature, density and magnetic field (van der Holst 2014; Oran et al. 2017).

Inspired by these applications to the Sun, *AWSoM* code was applied for stellar coronal and wind environments of cool stars by Alvarado-Gómez et al. (2016a,b). All these studies utilized observationally derived ZDI magnetograms as model input, but the Alfvén wave flux was assimunot constrained by observations of the simulated stars. Boro Saikia et al. (2020) also used this model to study the effect of low-resolution stellar magnetograms to simulate the wind mass loss rates from the Sun and a young solar-like star, HN Peg. The results suggest that the solar wind model based on synoptic magnetograms derived from solar Global Oscillation Network Group (GONG) magnetograms degraded to the low spherical harmonic number ($l_{max}$ = 5-10) do not significantly affect the solar wind mass loss rates as compared to the high resolution magnetograms, because the wind structure is not sensitive to small-scale magnetic field structures. However, the stellar coronal X-ray and EUV emission was not addressed in these studies.

Here, we present the results of data constrained fully thermodynamic data-constrained 3D MHD models of the solar corona and the wind at the rising phase of Solar Cycle 24 and the stellar corona of a young (~650 Myr) solar analog, $\kappa^1$ *Ceti* (HD 20630). The paper is organized as follows. Section 2 describes the data driven *AWSoM* models for the Sun and $\kappa^1$ *Ceti*. In Section 3, we present the results of MHD simulations of the global corona of $\kappa^1$ *Ceti* at two different epochs and describes the stellar wind and corotating interaction events (CIRs) from $\kappa^1$ *Ceti*. In Conclusion, we discuss the implications of this modeling methodology for various active stars and impact on the early Earth and terrestrial type exoplanets around active stars.

2. Data driven 3D MHD corona and wind model

To model the solar and stellar corona-wind system of $\kappa^1$ *Ceti*, we used a data-constrained Alfvén wave-driven 3D MHD numerical model, *AWSoM*. This is a first-principles global model that describes the stellar atmosphere from the top of the chromosphere and extends it into the heliosphere beyond Earth's orbit. *AWSoM* model is a part of the Space Weather Modeling

Framework (*SWMF*) developed at the University of Michigan. This framework provides a high-performance computational capability to simulate solar activity induced by the magnetic flux from the upper chromosphere to the solar wind and planetary environments (Toth et al. 2005, 2012; Jin et al. 2012; 2017; van der Holst et al. 2014; Oran et al. 2017).

The AWSoM model solves a set of two-temperature (electrons and protons) MHD equations for fully ionized plasma in the heliographic rotating frame by the shock-capturing MHD BATS-R-US code (Powell et al. 1999). The model assumes that the Hall effect is negligible and the electrons and protons have the same bulk velocity. Thus, the code uses a single-fluid continuity and momentum equations with separate pressure equations for electrons and protons. The *AWSoM* consists of two modules describing solar/stellar corona (SC) and inner helio/astrosphere (IH) respectively. The SC module uses a 3D spherical grid with the radial coordinate ranging from 1 $R_\odot$ to 24 $R_\odot$. The grid is highly stretched toward the star with smallest radial cell size $\Delta r = 10^{-3}$ $R_\odot$ to numerically resolve the steep density gradients in the upper chromosphere and transition region. The IH module describes the helio/astrosphere from 16 $R_\odot$ to 250 $R_\odot$, so that SC and IH overlap. Thus, to obtain the solution from the Sun to the Earth, we couple the SC and IH components.

The MHD equations are coupled to wave kinetic equations for propagating parallel and anti-parallel Alfvén waves (van der Holst et al. 2014). A steady-state solar wind solution is obtained with the local time stepping and second-order shock-capturing scheme. In this model, the low frequency (compared to the ion cyclotron frequency) torsional Alfvén waves launched at the upper chromosphere (lower computational boundary) drive the plasma dynamics by exchanging momentum and energy with the plasma: gradients in the wave pressure accelerate the plasma, while dissipation converts wave energy into thermal energy. The model includes two equations that describe the amplitudes of propagating low-frequency Alfvén waves parallel and anti-parallel to the magnetic field and are coupled to the MHD equations. Alfvén wave energy dissipation is the only explicit source of coronal heating incorporated into the energy equation, and no *ad hoc* or geometric heating functions are used in this model. This global coronal model does not resolve the wave motions, because the time and spatial scales associated with the wave processes are much smaller than the characteristic scales of the corona. This allows to treat the wave energy evolution under the WKB approximation. In our model, the coronal heating is driven by the magnetic energy propagated and dissipated via turbulent cascade of Alfvén waves in both closed and open field regions. As low-frequency Alfvén waves propagate upwards into a gravitationally stratified stellar atmosphere, they become subject to reflection from regions of high gradients of Alfvén velocity (Heinemann & Olbert 1980; An et al. 1990). The interaction of downward-reflected Alfvén waves with upward propagated ones can ignite a turbulent cascade of Alfvén waves in the lower solar atmosphere and provide a dominant source for heating the solar and stellar coronae and winds in open field regions (Cranmer & Ballegooigen 2005; Cranmer 2011; Sokolov et al. 2013). Within the closed loop-like magnetic structures the waves of different polarities (along and opposite to the direction of the loop magnetic field) can interact in the similar fashion producing the heating within the coronal active regions. AWSoM model provides the self-consistent evolution of the wave energy coupled to an MHD plasma of the solar corona (see van der Holst et al. 2014 for details). The wave energy in this model represents the time average of the perturbations due to a turbulent spectrum of Alfvén waves. Because energy dissipation of these waves controls the density and the temperature of the stellar corona, the accurate knowledge of the Poynting flux needs to be constrained from spectroscopic observations of stellar chromospheres.

The plasma density and the Alfvén wave amplitude can be derived from the Far UV optically thin spectral lines forming at about 50,000K specified by the *AWSoM* model. To model the solar corona, studies use IRIS hight spectral observations of fully resolved profiles of OIV and SiIV emission lines (Oran et al. 2017). The model inputs include high spatial resolution solar magnetograms and Alfvén wave energy flux specified at the inner boundary, the upper chomosphere at 50,000 K as the wave Poynting flux normalized to the magnetic field, $S_A/B$. This parameter can be derived directly from NASA's Interface Region Imaging Spectrograph (*IRIS*) of fully resolved spectral emission lines of OIV and SiIV ions forming in the upper atmosphere (Sokolov et al. 2013; Young et al. 2018). In stellar wind models $S_A/B$ input Alfvén wave energy flux is not directly constrained from observations, but is scalled with the stellar X-ray flux and the surface magnetic flux (Pevtsov et al. 2003; Garraffo et al. 2016; Dong et al. 2018). However, this requires prior information of the star's X-ray activity. As stellar X-ray activity is known to exhibit variations, this approach will also lead to variations in $S_A/B$ for a magnetically variable star. In the solar case the Alfvén wave Poynting flux is well constrained from observations and nearly constant for all solar simulations. The value of $S_A/B$ is a critical input parameter, and the wind mass flux is roughly proportional to the solar/stellar wave Poynting flux (Cranmer & Saar 2011, Boro Saikia et al. 2020).

3. Simulation Results

To model the global solar and stellar corona-wind system of $\kappa^1$ *Ceti*, we used a data-constrained Alfvén wave-driven, global MHD numerical model, *AWSoM*. The global parameters for these two stars are presented in Table 1. In Section 3.1, we present the model inputs and steady state solutions for the solar coronal structure and the solar wind simulated for the Carrington Rotation 2106, which is representative of the early rising phase of solar cycle 24. For both models, we calculated thermodynamic parameters including temperature, density, magnetic field and velocity of global solar corona. These parameters are used to build synthetic coronal emission maps in soft X-ray band (0.25-4 keV) and Extreme UV lines including at 284Å, 335Å and the soft X-ray 1-8 Å band (referred to as SXR) to derive the benchmark solar model for the stellar coronal model of $\kappa^1$ *Ceti*. In Section 3.2, we present the results of the 3D MHD modeling of the corona of $\kappa^1$ *Ceti* in SXR and EUV bands and discuss the structures of the associated stellar wind at two epochs, 2012.8 and 2013.7.

Table 1. Observational properties of current and the young Sun represented by $\kappa^1$ *Ceti*

| Star ID | Sp Type | $T_{eff}$(K) | Mass | Log (g) | $P_{rot}$ (d) | i (deg) | Age (Myr) | Log ($L_x$) |
|---|---|---|---|---|---|---|---|---|
| $\kappa^1$ *Ceti* | G5V | 5705 | 1.02 | 4.49 | 9.2 | 60 | 650 | 28.79 |
| Sun | G5V | 5780 | 1 | 4.5 | 27 | 0 | 4650 | 27 |

3.1 Global Coronal Model of The Sun: Carrington Rotation 2106

To compare the current Sun's coronal and wind fluxes with the corresponding fluxes from $\kappa^1$ *Ceti*, we perfomed the model of global solar corona with *AWSoM* at the phase near solar minimum for Carrington Rotation, CR 2106. This corresponds to the rising phase of Solar Cycle 24. The inner boundary condition of the solar surface magnetic field is specified by a global magnetic map sampled from an evolving photospheric flux transport model for the periof of Jan 20-Feb 11, 2011) (Schrijver & DeRosa 2003). The input Alfvén wave Poynting flux normalized to the magnetic field is specified at the lower boundary of the Sun, the upper chromosphere and described as

$$P_W = \frac{S_A}{B_0} = \sqrt{\frac{\rho}{4\pi}} < \delta V^2 >, \qquad (1)$$

The solar coronal model uses the input plasma density of $2 \times 10^{11}$ cm$^{-3}$ at 50,000K and the Alfvén wave amplitude, $(< \delta V^2 >)^{1/2} = 15$ km/s derived from OIV and Si IV spectral lines derived from IRIS data (see discussion in Sokolov et al. 2013; der Holst et al. 2014; Oran et al. (2017). These parameters output the Poynting flux, $P_{w0} = (S_A/B_0)_\odot = 1.1 \times 10^5$ erg/cm$^2$/G . We will use this reference normalized solar chromospheric Alfvén wave Poynting flux, for our stellar coronal model discussed in Section 3.2. The adjustable input paramaters include the transverse correlation length at the inner boundary, $L_\perp$ and the pseudoreflection coefficient, $C_{refl}$ as discussed in details by Oran et al. (2013).

We have performed the 3D MHD simulations of the solar corona until the solution is converged to steady state corona and the wind solution. We then used the 3D model density and temperature simulated values to synthesize coronal emission maps at Fe XV 284A and Fe XVII 335A EUV emission lines and GOES 1-8Å soft X-ray flux by integrating along the line of sight (*LOS*) towards the observer using CHIANTI 7.1 atomic database for solar abundances (Dere et al. 1997). The total intensities integrated over the solar disk at 1 AU in the soft X-ray 1-8Å band is $2.7 \times 10^{-5}$ erg/cm$^2$/s , while the intensities in EUV lines, Fe XV 284Å and Fe XVII 335Å coronal lines are 0.014 erg/cm$^2$/s and 0.005 erg/cm$^2$/s, respectively. These fluxes and GOES (1-8Å) band fluxes apper to be consistent with observed values in the rising phase of SC 24 (see for example, Winter and Balasubramanian 2014; Huang et al. 2016). The simulated mass loss rate of the Sun during CR 2106 is $1.8 \times 10^{12}$ g/s (or $2.86 \times 10^{-14}$ $M_\odot$/yr) and the minimum and maximum dynamic pressures of the solar wind, $\rho V_w^2$ , over the Earth's orbit are 0.7 and 6.6 nPa, respectively. These values are consistent with *WIND* observations of the solar wind at 1 AU (Mishra et al. 2019). The MHD model outputs including the SXR band and EUV coronal line intensities and solar wind mass loss rate are presented in Table 1.

*3.2* Global Coronal Model of $\kappa^1$ *Ceti*

$\kappa^1$ *Ceti* is a young, 750 Myr old, nearby (9.16$\pm$ 0.06 pc) G5V dwarf. Its age of 750 Myr was estimated by (Güdel et al. 1997) from the rotation rate of 9.2 d, but later age estimates suggest earlier age of ~600 Myr (Ribas et al. 2010), which makes this star one of the best proxies of the young Sun at the time when life started on Earth (Cnossen et al. 2007; Ribas et al. 2010; Do Nascimento et al. 2016). $\kappa^1$ *Ceti* was a subject of a number of comprehensive multi-wavelength studies that provided the measurements of its surface magnetic field, chromosphere, transition region and coronal fluxes, surface spot and flare activity (Güdel et al. 1997; Messina & Guinan 2002; 2003; Teleschi et al. 2005; Ribas et al. 2005; Schaefer et al. 2000; do Nascimento et al. 2016; Rosén et al. 2016). The star shows the signatures of magnetic activity in the form of intense X-ray coronal flux enhanced by a factor of 40 with respect to the solar flux at solar maximum, energeric flares and and the presence of large starspots covering from 1% to 9% of the stellar hemisphere (Schaefer et al. 2000; Walter et al. 2007). To model the global stellar corona-wind system of $\kappa^1$ Ceti, we used the *AWSoM* model with the input stellar magnetogram derived from spectropolarimetric observations of the star at two epochs as discussed in Section 3.2.1 and Alfvén wave flux constrained from the HST/STIS observations of the star presented in Secttion 3.2.2.

### 3.2.1 Magnetic map of $\kappa^1$ Ceti

The stellar magnetic field can be specified by a photospheric magnetic map derived from spectropolarimetric observations. The unsigned magnetic field averaged over the stellar surface, *fB*, can be directly derived from unpolarized (Stokes I) observations using the Zeeman broadening technique (Robinson et al. 1980; Reiners & Basri 2006; Kochukhov et al. 2020). Global stellar photospheric magnetic field can also be reconstructed by inverting time series of high-resolution spectropolarimetric data (circularly polarized Stokes V profiles) with the procedure known as Zeeman Doppler Imaging (ZDI) technique (Semel 1989). Here we used the surface magnetic field maps of $\kappa^1$ Ceti (see Figure 1) at two different epochs, 2012.8 and 2013.7. These maps were reconstructed by Rosén et al. (2016) from the time-series Stokes V observations with the spectropolarimeter NARVAL at the 2m Telescope Bernard Lyot at the Pic du Midi Observatory (France) and the spectropolarimeter HARPSpol at the ESO 3.6m telescope in La Silla, Chile.

The ZDI analysis proceeded as follows. First, the signal from a large set of spectral lines was combined in order to obtain a set of high S/N mean line profiles, which is necessary for detecting weak polarization signatures. This multi-line analysis was accomplished by applying the least squares deconvolution technique (LSD, Donati et al. 1997; Kochukhov et al. 2010) using a line list retrieved from the VALD3 database (Ryabchikova et al. 2015) for a MARCS stellar atmosphere (Gustafsson et al. 2008) with $T_{\text{eff}}$ = 5750 K and log g = 4.4. Only lines stronger than 20% of the continuum were used and spectral regions contaminated with tellurics or dominated by particularly strong and broad spectral lines were removed. The LSD Stokes *IV* profiles were then computed with the LSD code described by Kochukhov et al. (2010). The resulting circular polarization profiles for September of 2012 (2012.8) and August of 2013 (2013.7) epochs of observations of $\kappa^1$ Ceti are shown in the left panel of Fig. 1.

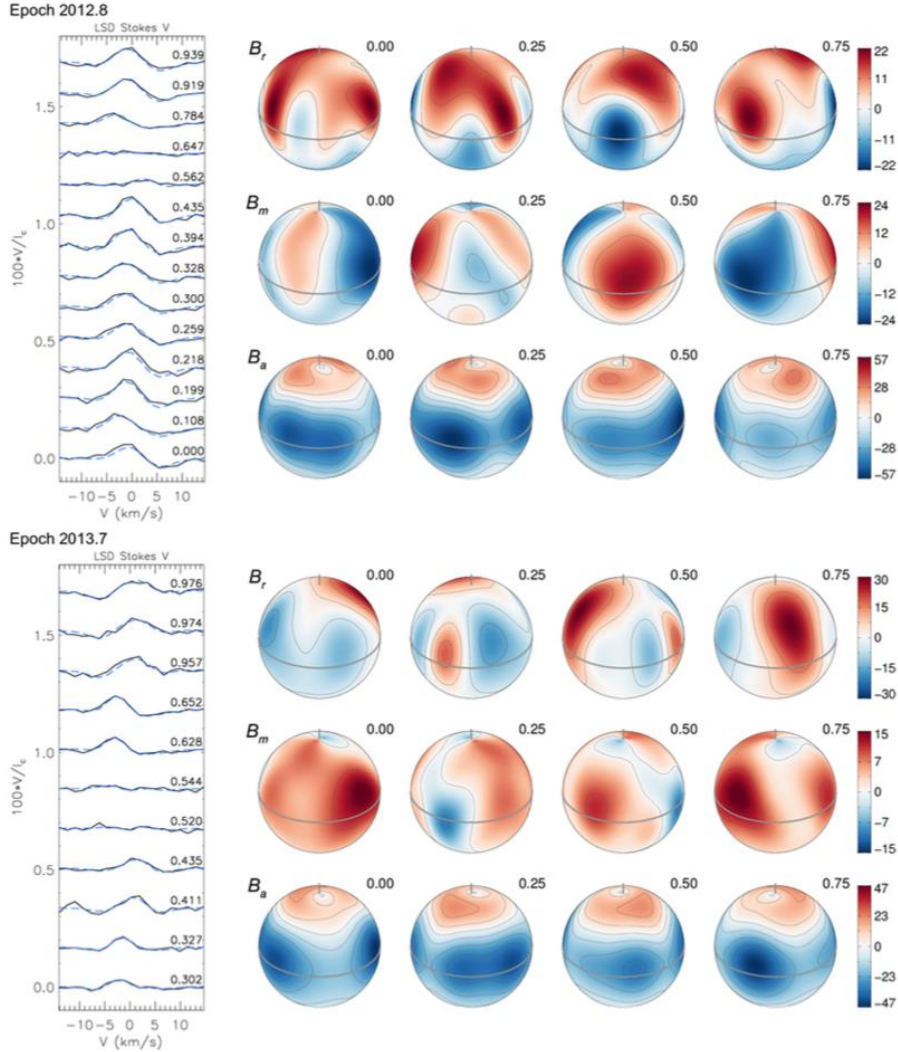

*Figure 1. LSD Stokes V profiles (left panel) and the reconstructed global radial, $B_r$, azimuthal, $B_\phi$, and meridional $B_\theta$ field components (right panel) of $\kappa^1$ Ceti for the epochs of 2012.8 and 2013.7 respectively shown for four rotation phases (*Rosén et al. 2016*).*

To reconstruct the large-scale photospheric magnetic field geometry of $\kappa^1$ Ceti, Rosén et al. (2016) used the Zeeman Doppler imaging code InversLSD (Kochukhov et al. 2014). They assumed the projected rotational velocity of 5 km/s (Valenti & Fischer 2005) and an inclination angle of 60°. The stellar magnetic field geometry was parametrized using a spherical harmonic expansion truncated at $l_{max} = 5$. The resulting maps of the radial, $B_r$, azimuthal, $B_\phi$, and meridional $B_\theta$ field components are presented in the right panel of Figure 1 for the epochs of 2012.8 and 2013.7 respectively. The maximum angular degree assumed in the ZDI analysis corresponds to the large-scale field with resolved scales $> 180^0/l_{max}$ (Morin et al. 2010; Johnstone et al. 2014), and thus misses about 90% of the magnetic flux associated with small-scale structures including star spots and due to flux cancellation effects (See et al. 2019). Indeed, for $\kappa^1$ Ceti, the unsigned magnetic field varies between 450 and 550 G (Kochukhov et al. 2020) and is by a factor of ~20 greater that the global field derived from these ZDI magnetograms, which is also consistent with simulations for active solar-type stars (Lehmann et al. 2019). Thus, this model can only account for the magnetic flux fro the large-scale structures, this represents the low bound of the surface magnetic flux of this star.

The surface magnetic field was then extrapolated to a 3D potential field source surface (PFSS; Wang & Sheeley 1992) solution using the finite difference iterative potential- field solver (FDIPS). The PFSS model assumes the potential (current-free) magnetic field ($\nabla \times B = 0$) that satisfies Laplace's equation ($\nabla^2 \Phi = 0$). As an upper boundary condition, we assume that the magnetic field becomes radial at a *source surface* which is taken to be at a height of 2.5 $R_\odot$. The FDIPS method avoids the ringing patterns near regions of concentrated magnetic fields to which the spherical harmonics method is susceptible, see Toth et al. (2011). Figure 1 shows that magnetic field components changes dramatically in geometry and magnitude over 11 months of observations. The upper panel of Figure 1 suggests that the global magnetic field is mostly poloidal and resembles the global field of our Sun during solar minimum, while 11 months later it became tilted at 45 degrees with respect to the ecliptic plane with the presence of toroidal component to be discussed in Section 3.2.3. Such variability of global magnetic field geometry on a time scale of a few months to one year is a typical signature of young solar-like stars that is traced by ZDI magnetograms of young solar type stars like BE Ceti, HN Peg and resulting X-ray coronal emission fluxes (Gudel 2007; Boro Saikia et al. 2015; Rosen et al. 2016). It is interesting to mention that a young, 300 Myr old solar-like star, $\chi^1 Ori$, also shows a short (a few months) variability in the large-scale magnetic field structure (Waite et al. 2017).

3.2.2 The Input Alfvén wave Poynting flux for $\kappa^1$ Ceti

At the lower computational boundary of the *AWSoM* model, we specify the Poynting flux of Alfvén waves normalized to the local magnetic field, $P_w = S_A/B_0$, as

$$P_{AW} = \sqrt{\frac{\rho}{4\pi}} < \delta V^2 >, \qquad (1)$$

where $\rho$ is the plasma mass density, $< \delta V^2 >$ is the square of the Alfvén wave amplitude averaged over the time scale greater that the wave period (Sokolov et al 2013). In (1), the incompressible Alfvén wave density and the velocity fluctuations vary independently in the chromosphere as the waves are weakly non-linear. As waves propagate upward, their amplitude increases with height, and as Alfvén waves become strongly non-liner, they excite longitudinal waves (Sabri et al. 2020)

To obtain the normalized Poynting at the upper chromosphere of $\kappa^1$ Ceti, $(S_A/B_0)_{star}$, we derived the input chromospheric plasma density of the star from the HST/STIS Far UV observed (E140M mode) fully resolved spectral lines OIV 1399.78, 1401 and SiIV 1401, 1393 A ions in spectra of $\kappa^1$ Ceti presented in Figure 2.

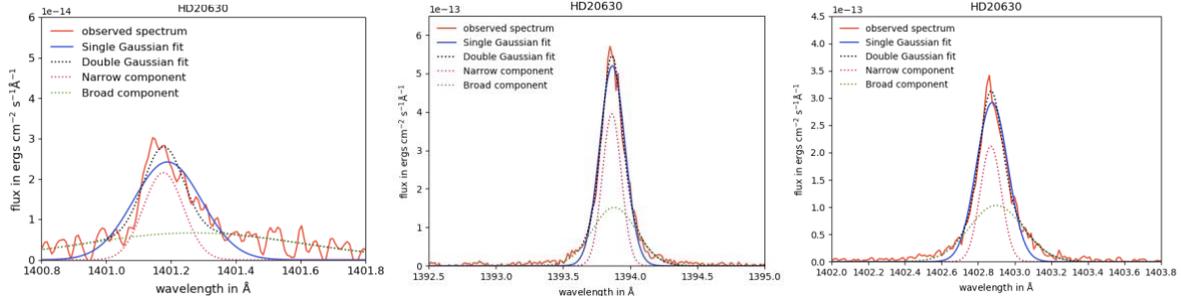

*Figure 2. The profi* MHD Poynting flux input plasma density of $10^{13}$ cm$^{-3}$. According to the Vernazza-Avrett-Loeser (VAL- C) model presented in Vernazza et al. (1981), the upper chromospheric density at the layers of formation of SiIV and OIV emission lines does not vary more than a factor of 2 (average quiet Sun to very bright network element). Thus, this uncertainty in the chromospheric density will affect the Poynting flux by about 40%. Also, according to Wedemeyer-Böhm et al. (2009), the structure of the upper chromosphere can be represented as a canopy of horizontal internetwork magnetic fields rather than well-defined open and close magnetic flux characteristic of the solar and stellar coronae. To refine the inputs to future data constrained solar and stellar MHD coronal models, detailed characterization of chromospheric flux variation inside and outside solar and stellar active regions is required.

The amplitude, $<\delta V^2>$, of low frequency unresolved transverse Alfvén waves can be derived from the measured FWHM of OIV and SiIV emission lines. These lines show non-thermally broadened profiles caused by the unresolved large-scale turbulent (non-thermal) motions of OIV and SiIV ions. In the general case, where the non-thermal motions are assumed to be random, the observed/measured FWHM of an optically thin emission line are given by (Phillips et al. 2008)

$$FWHM = \sqrt{\Delta\lambda_{inst}^2 + 4ln2\left(\frac{\lambda_0}{c}\right)^2\left(\frac{2k_BT_i}{M_i} + V_{turb}^2\right)} \quad (2)$$

where $\Delta\lambda_{inst}$ is the instrumental broadening, $\lambda_0$ is the rest wavelength, c is the speed of light, $k_B$ is the Boltzmann constant, $T_i$ and $M_i$ are the temperature and atomic mass of ion i, respectively, and $V_{turb}$ is the non-thermal ion speed along the line of sight. Here, we assume that the non-thermal motions of coronal ions are caused by the Alfvén waves, which cause the ions to move with a velocity equal to the rms wave velocity perturbation, $\sqrt{<\delta V^2>}$. Then, the rms wave amplitude can be derived directly from the turbulent velocity as

$$\sqrt{<\delta V^2>} = \frac{2V_{turb}}{|cos\alpha|},$$

where α is the angle between the plane perpendicular to the magnetic field and the line-of-sight vector (Hassler et al. 1990). To derive the initial amplitude of Alfvén waves at the lower boundary, we fitted the fully resolved observed profile of OIV and SiIV lines (see Figure 2) obtained with high-resolution HST/STIS observations. FWHMs is derived from the narrow component of Gaussian line fits. We then subtracted the Doppler and instrumental broadening components to obtain the non-thermal broadening according to Eq. 2. This results in the turbulent velocity of 24.6 km/s, which is ~ 1.6 times greater than that measured in OIV lines of the current Sun (Sokolov et al. 2013). Assigning the non-thermal line broadening to large-scale turbulent motions via unresolved Alfvén waves, we derived the average wave amplitude, $V_w = \sqrt{<\delta V>^2}$. Thus, in our models stellar we combined the density derived from the line ratios with this turbulent velocity to models for 2012.8 (M2012) and for 2013.7 (M2013) epochs, we set the input normalized Poynting flux, $P_w = 26.9\ P_{w0}$, where $P_{w0} = 1.1 \times 10^5$ erg/cm$^2$/G is the reference normalized solar chromospheric Alfvén wave Poynting flux, used in our current and earlier solar modeling studies (Sokolov et al. 2013; van der Holst et al. 2014; Oran et al. 2017).

### 3.2.3 The Coronal Structure and Emission from $\kappa^1$ Ceti

We calculated coronal models of $\kappa^1$ Ceti using the reconstructed magnetogram for 2012.8 epoch referred to as StellarE1 and 2013.7 epoch referred to as StellarE2 for respectively. The left panel of *Figure 3* shows that the converged (steady) solution for the global magnetic coronal structure of the star at 2012.8 epoch, which is mostly represented by a dominant dipolar field tilted at $9^0$ (dipole strength of 15.38 G vs 5.1 G and 7.99 G for quadrupole and octupole respectively at 2013.7) and resembles the current Sun's coronal state at minimum of the solar cycle. The right panel demonstrates the magnetic filed topology 11 months later suggesting that the stellar dipole magnetic field has undergone a dramatic transition from a simple dipole to $45^0$ tilted dipolar magnetic field with over 2/3 of the contribution from quadrupolar and octupal components). This configuration is typical for the current Sun's magnetic field at the declining phase of the solar maximum. The total unsigned open magnetic flux at 2012.8 epoch is $6.2 \times 10^{23}$ Mx, while the flux decreases to $4.5 \times 10^{23}$ Mx at 2013.7 epoch.

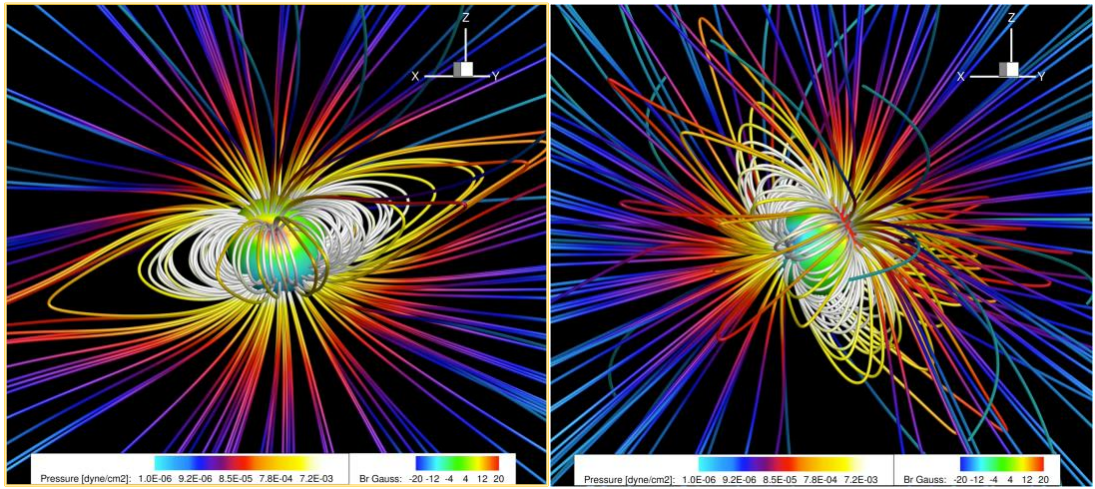

*Figure 3. Global structure of the stellar magnetic corona of $\kappa^1$ Ceti at 2012.9 (left panel) and 2013.8 (right panel) with the superimposed plasma pressure along the field lines specified by the color bars.*

*Table 2. The summary of the stellar coronal intensities and wind mass fluxes from the solar model M0, StellarE1 and StellarE2 models.*

| Star | FeXV 284A erg/cm$^2$/s | FeXVII 335A erg/cm$^2$/s | SXR 1-8A in 10$^{-5}$ erg/cm$^2$/s | $\dot{M}$ in $10^{-14} M_\odot$/yr | V$_{fast}$/V$_{slow}$ km/s | P$_{dyn}$/ P$_E$ |
|---|---|---|---|---|---|---|
| Sun/CR2106 (M0) | 0.014 | 0.005 | 2.7 | 2.86 | 700/450 | 0.7-6.6* |
| $\kappa^1$ Cet - 2012.8 StellarE1 | 0.42 | 0.22 | 7000 | 286 | 1152/696 | 188-1379* |
| $\kappa^1$ Cet - 2013.7 StellarE2 | 0.27 | 0.16 | 4500 | 238 | 1253/692 | 104-958* |

*Dynamic pressure of the stellar wind is measured at 1 AU with the minimum and maximum values around the Earth's orbit.

The calculated output SXR fluxes in Fe XV 284Å, Fe XVII 335Å lines and the Soft X-ray (SXR) band 1-8Å from the stellar StellarE1 and Stellar E2 models are presented in Table 2. The integrated fluxes in Fe XV 284Å, Fe XVII 335Å emission lines during 2012.8 epoch are by 50% and 37% larger than that at 2013.7. These simulated fluxes are by a factor of 5-10 smaller than measured from the EUVE observations of this star in 1995 (Ribas et al. 2005). The major reason of the underestimated flux is in the incomplete representation of the surface magnetic flux via ZDI reconstruction. As we discussed in Section 3.2.1, the stellar magnetic map geometry can only represent the large-scale magnetic structures $> 72^0$, the maximum spherical harmonic expansion is $l_{max}$ = 5. Such low spherical harmonic number representation of the global magnetic field misses over 10 times of the total unsigned magnetic field (See et al. 2019). Indeed, measurements of $\kappa^1$ Ceti unsigned magnetic flux from Zeeman broadening of unpolarized spectra suggest the average magnetic field ~ 500 G (Kochukhov et al. 2020), which is about 20 times greater than that inferred from ZDI maps presented in Figure 1. This is the result of an unresolved flux in Stokes V (circularly polarized) observations due to magnetic flux cancelation within pixel resolution or starspots due to suppression of the Zeeman effect in dark regions (flux cancellation of oppositely signed spots). Thus, about 95% of the magnetic flux is concentrated in smaller magnetic structures mostly represented by bipolar regions or starspot groups. To resolve such small scale structures, high precision photometric techniques such as transit observations by Kepler, K2, TESS or other photometry like Evryscope, Howard et al. 2020) are required. In the transit observations, starspots are usually identified as rotationally modulated "dips" that appear as dark regions on the stellar disk (Namekata et al. 2019). *MOST* observations of $\kappa^1$ Ceti taken in 2003-2005 suggest that the surface magnetic flux is mostly concentrated in two starspots in 2003 (with areas of 3.6 and 1%), three spots in 2004 (with areas of 1.9%, 9% and 5.3%) and two spots in 2005 covering 2.9% and 2.2% of the stellar disk respectively (Rucinski et al. 2004; Walker et al. 2007). These observations suggest that the lifetime of large spots do not exceed one year and are consistent with the recent estimates of such large starspots from young solar-type stars in Kepler data (Namekata et al. 2019). This suggests that most of the magnetic flux is concentrated at angular sizes of $< 16^0$, and thus would require the ZDI reconstruction with $l_{max} > 15$.

In order to reconcile to the total average unsigned magnetic flux measured in unpolarized observations (Kochukhov et al. 2020), we will use the *MOST* mission 2003-2005 data and our recent *TESS Cycle 1* data of the rotational modulation of the star to characterize the starspot(s) to be added as bipoles to the ZDI magnetograms. These localized structures carry over 90% of the unsigned magnetic flux that should, therefore, provide the dominant contribution of the coronal SXR and EUV fluxes as it follows from empirical nearly linear relationship, $F_{SXR} \propto \Phi^{1.15}$, between the stellar SXR and surface magnetic fluxes (Pevtsov et al. 2003). The estimates of the expected contribution of starspots to the overall EUV and SXR flux can be based "Sun-as-a-star" study of a single sunspot representd by the active region (AR 12699) transited throughout the solar disk (Toriumi et al. 2020). This transiting sunspot with the area, A=240 MSH (Millionth of the Solar Hemisphere), was associated with > 10% enhancement of the coronal emission in the EUV FeXV 284Å and FeXVII 335Å emission lines and SXR 1-8A flux by a factor of ~10 times of the quiet Sun's flux. We can scale the spot's contribution to the stellar coronal emission using the *κ¹ Ceti*'s starspot areas observed with MOST from 10,000-90,000 MSH. Assuming the starspot's magnetic field to be comparable to the sunspot's field, we can estimate the lower bound of the magnetic flux of a starspot asscociated with the coronal active region, $\Phi = B_0 A$, where $B_0$ is the starspot's magnetic field strength, to be $\geq$ 40-375 times of the AR 12699 flux. Then, using the the solar/stellar SXR and surface magnetic fluxes, we find that the low limit of the expected contribution to the EUV and SXR stellar flux from the starspot should be > 7-90 times of the quiescent (unspotted) area of the star modelled here. We will study these contributions using the MHD model of the star in the upcoming study.

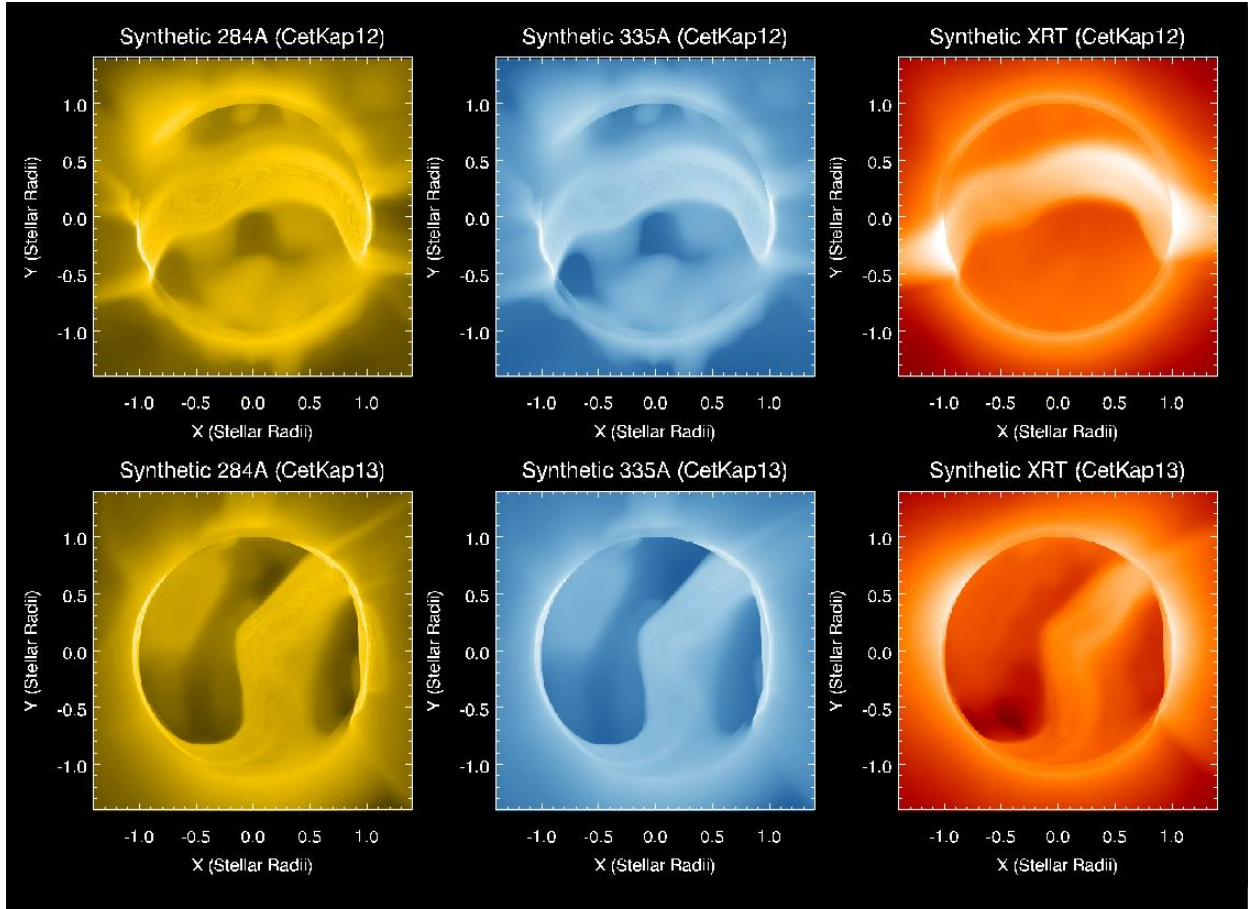

*Figure 4. The coronal images of the Sun (upper raw and at 2012.9) and 2013.8 epoch (low raw) at 284A, 335A and XRT band.*

We used CHIANTI code to construct the 3D synthetic coronal emission maps of $\kappa^1$ *Ceti* at two epochs, August 2012 (2012.8) and July of 2013 (2013.7). The 2D slices of the 3D coronal images of $\kappa^1$ *Ceti* are presented in Figure 4. The upper panel of the figure shows the X-ray and EUV emiting regions of the corona at 2012.8 of the star in the mid- to higher norther latitudes and the well pronounced coronal hole formed around the southern pole of the star. However, the coronal flux becomes significantly different 11 months later as the magnetic dipole flips, exposing the bright X-ray-EUV emitting corona with 40% less flux due to stronger disorganized field (less closed magnetic flux) in the southern to mid-lattitudes of the star and the formation of the coronal hole appearing in the northern regions of the star. We used these data to calculate the volumetric emission measure, VEM=$n_e n_i V$ ($n_e$, $n_i$ are the electron and ion number densities, V is the volume of emitting plasma) distribution over temperatures. *Figure 5* shows the simulated VEM distribution of the solar corona (black line), the stellar corona at 2012.8 (red line) and 2013.7 (blue line) with the dominant contribution at 6.3 MK, which is consistent with the XMM-Newton spectra of the star (Teleschi et al. 2005). It is evident that the weaking of the dipole field and appearance of multipole components reduced the VEM at this temperature by a factor of 2.5 with the maximum temperature at ~ 6 MK.

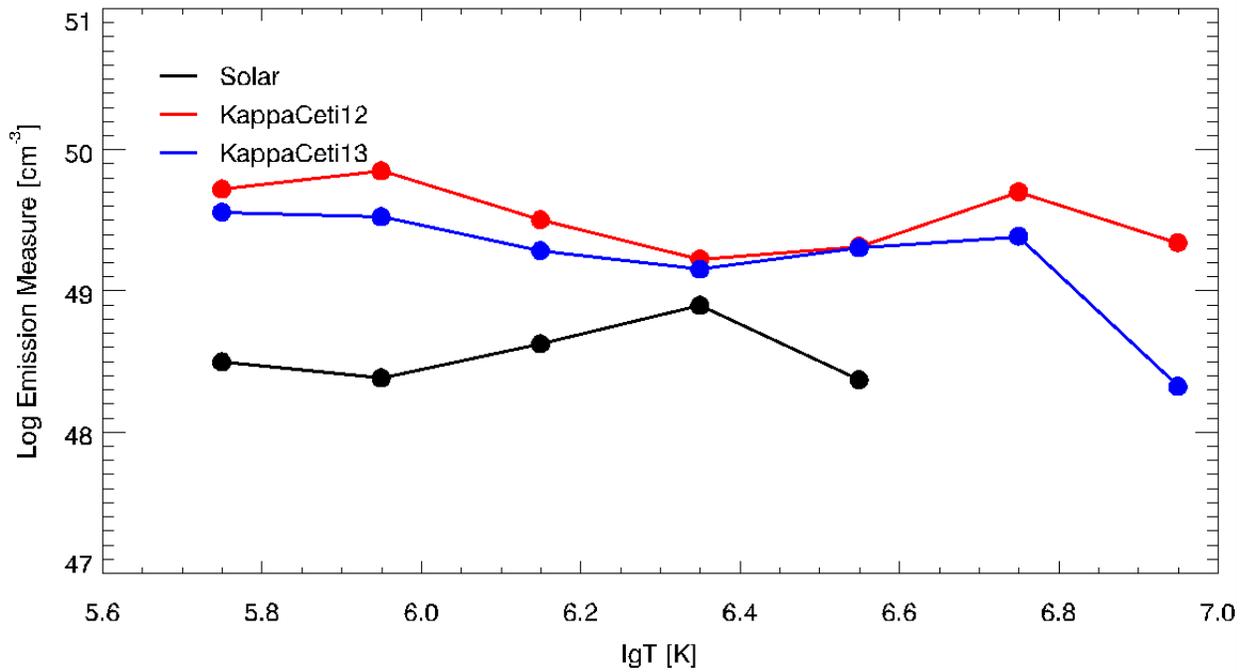

*Figure 5. Volumetric Emission Measure distribution over temperature for the Sun (black), κ¹ Ceti at 2012.8 (red) and 2013.7 epochs (blue).*

3.2.4 The stellar wind from $\kappa^1$ Ceti

A. Mass Loss Rates.

Our *AWSoM* model produces steady wind solutions for 2012.8 and 2013.7 epochs. The global wind shows the two-component stellar wind of $\kappa^1$ Ceti. The plasma accelerated from giant coronal holes (see Figure 4) is driven by the Alfvén wave ponderomotive force, while slow wind is formed due to the thermal pressure gradient. At 2012.8 epoch, the fast wind reaches its terminal velocity of 1152 km/s within the first 15 $R_{star}$, while the slow and dense component of the wind originates from the regions associated with the equatorial streamer belt structures at 696 km/s. Once steady solution for the coronal density, magnetic field and velocity is obtained, we calculated the mass loss rate as

$$\dot{M} = \int_A \varrho\ v \cdot dA.$$

The simulated stellar wind density is ~50-100 times greater than that of the current Sun's wind and faster by a factor of 2, which produces a massive wind with the mass loss rates of 2.8 x $10^{-12} M_\odot$, which is 100 times greater than that observed from the current Sun. Our wind simulations are consistent with the empirically derived scaling of the mass loss rates with the SXR surface fluxes, $\dot{M} \propto F_X^{1.34}$, that provides the the upper bound of mass loss rate of a young G-type star with the surface X-ray flux of $10^6$ ergs/cm$^2$/s that is representative of $\kappa^1$ Ceti (Woods et al. 2014). However, this rate is twice larger than that obtained in the polytopic solutions of Airapetian & Usmanov (2016) and Do Nacsimento et al. (2016). We also find that the mass loss rate responds to the change of the large scale magnetic structures at 2013.7 epoch with 17% less mass loss rate. The resulting 2D (X-Y at Z=0, the ecliptic plane) maps of the wind density and dynamic pressure, $\rho V_w^2$ from $1R_\odot$ to $24R_\odot$ at two epoch in StellarE1 model is presented in Figure 5.

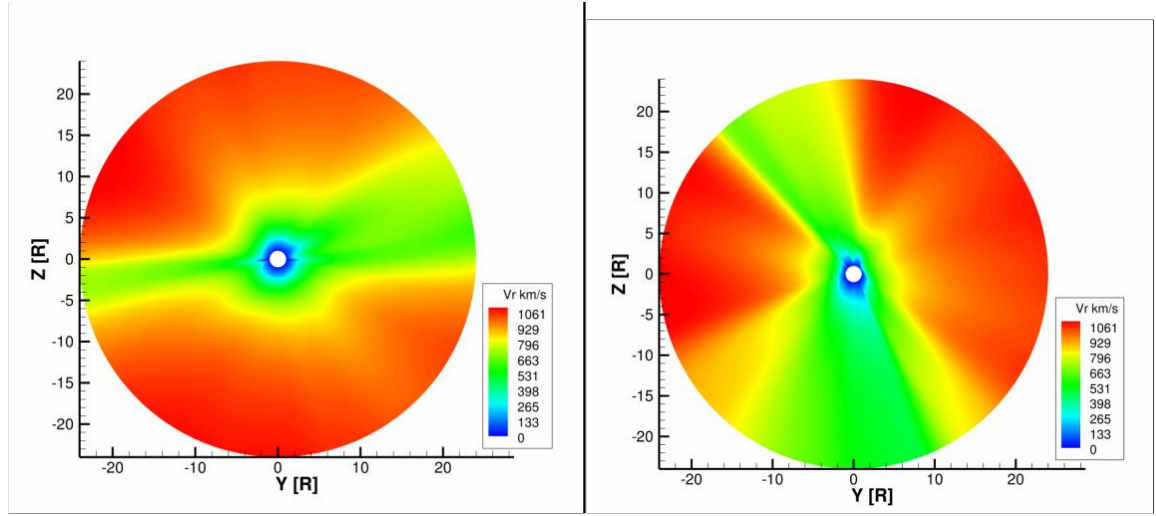

*Figure 6. The 2D Y-Z maps (at the ecliptic plane, X=0) of wind velocity in km/s (right panel), the dynamic pressure of the stellar wind in dyn/cm² over the first $24R_\odot$ (middle panel) and the dynamic pressure from the CIR structures from $1R_\odot$ to 1 AU from $\kappa^1$ Ceti formed by at 2012.8 (left panel) and 2012.7 (upper figure) and 2013.7 (lower figure) epochs. Black dashed lines show the orbit of Mercury, Venus and Earth, respectively.*

Our earlier sensitivity studies have shown that the wind mass loss rate increases about linearly with the with the input Alfvén wave Poynting flux (Boro Saikia et al. 2020). Such linear scaling can be explained due to the enhanced wave flux is dissipated increasing the coronal temperature, and thus causing larger acceleration of slow wind, while larger wave flux in the upper chromosphere drives faster winds in magnetically open regions drives due to the magnetic pressure gradient of waves.

B. Corrotating Interaction Regions.

Our steady solutions for $\kappa^1$ Ceti also show (see left and right panels of Figure 7) the formation of well pronounced regions of enhanced density of the stellar wind at two epochs, 2012.8 and 2013.7. These structures are the stelar wind stream interaction regions (SIR) formed due to the interaction of a stream of high-speed solar wind originating in the stellar coronal hole structures (see Figure 4) with the preceding slower wind formed along the equatorial regions of the stellar corona. The interaction of these steams forms a region of compressed plasma, SIR, along the leading edge of the stream, which, due to the rotation of the star at 9.2 days, is twisted approximately into an Archimedean (or Parker, 1958) spiral. Because the coronal holes may persist for many months, the interaction regions and high-speed streams tend to sweep past an exoplanet at regular intervals of approximately the half of the stellar rotation period (∼ 4.6 days) forming Corrotating Interaction Regions (CIRs) along the Parker spirals. These regions are signified by the compression region ahead of the fast wind and rarefaction region behind it that remain stationary in a frame corotating with the Sun shown in Figure 7. Due to much larger difference between the fast and slow wind components than that of the current Sun, its compression regions steepen into strong shocks propagating at > 1000 km/s at orbital locations of Mercury (0.4 AU), Venus (0.7AU) and Earth (1 AU).

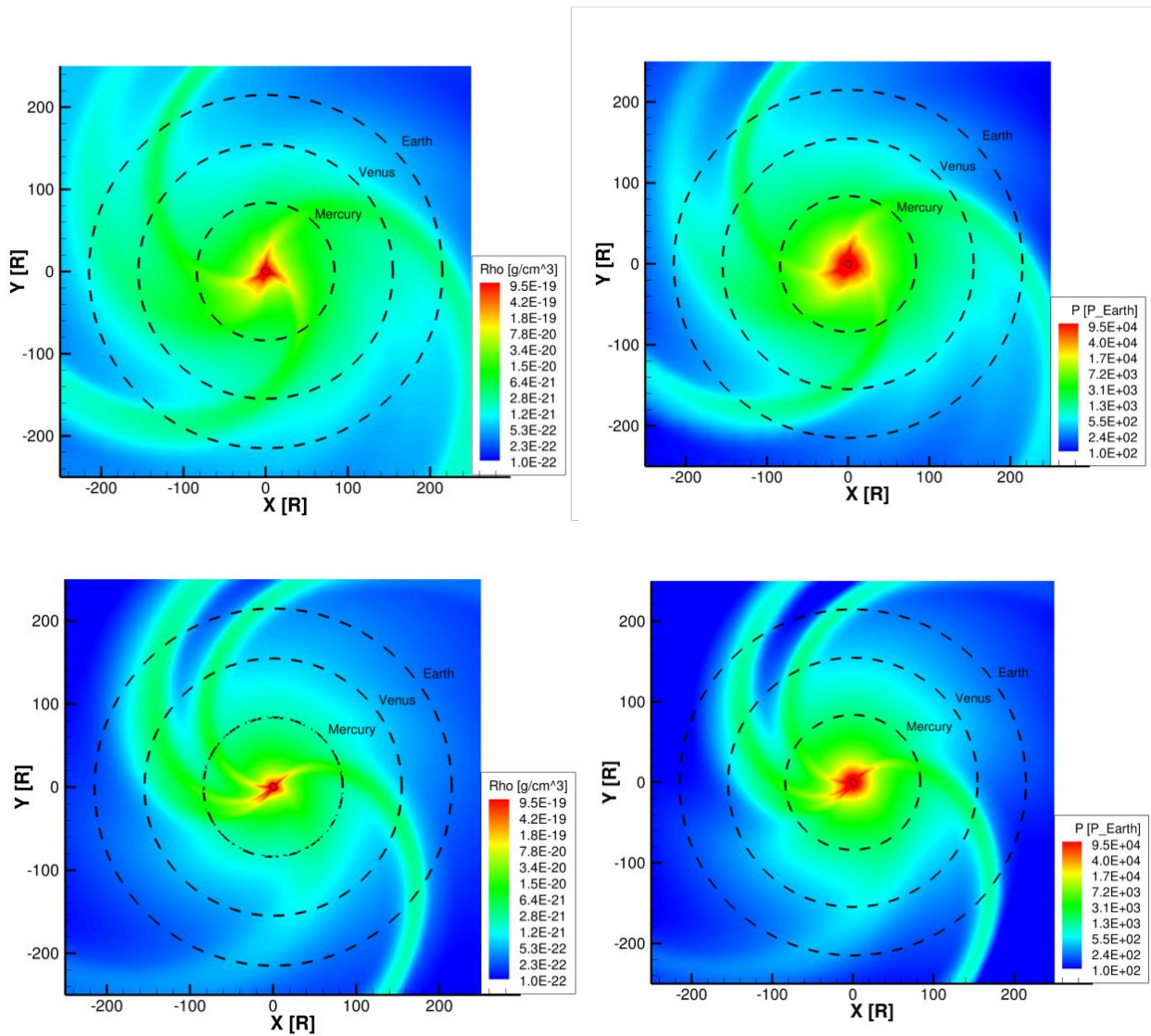

*Figure 7. The top panel: 2D slice of the X-Y plane (Z=0) of the plasma density and the dynamic pressure of CIRs from $\kappa^1$ Ceti formed at 2012.8 epoch (left) and 2013.7 epoch (right). Black dashed lines show the orbits of Mercury, Venus and Earth respectively.*

Figure 7 shows that stellar CIRs plasma density at $10R_\odot$ (0.05 AU) and $216R_\odot$ (1AU) are 5 x $10^5$ cm$^{-3}$ and 2 x $10^3$ cm$^{-3}$, respectively. These stellar wind densities are by a factor of 300 greater than the solar wind's density measured by Parker Solar Probe (Kim et al. 2020). These results are important as they provide the quantitative characterization of physical conditions at orbital distances of close-in exoplanets around G and K -types stars including eps Erib (Benedict et al. 2016), K2-229b (Santerne et al. 2018), K2-198b.c, K2-168b,c (Hedges et al. 2016), HD 189733b (Barth et al. 2021). The simulated dynamic pressure exerted by CIRs at close-in exoplanets reaches 50,000-10,000 times at 0.05 AU and 1300 times greater at 1 AU than that exerted to the magnetosphere of the current Earth. The stand-off distance of the stellar wind to the planet, $R_{sub}$, can be found from the balance between the dynamic pressure, $P_d = \rho v^2$, of the stellar wind and the magnetospheric pressure, $P_m$, of the dipolar planetary field as (Beard 1960)

$$R_{sub} \propto P_d^{1/6}.$$

This relation suggests that the dynamic pressure from stellar CIRs would compress the magnetospheres of Earth-like exoplanets (at the current Earth's magnetic moment) to the standoff distance of ~ $3R_\oplus$, and thus ignite strong geomagnetic currents in the early Earth atmosphere. Its interesting that Carrington-type CMEs modelled in Airapetian et al. (2016) can provide the comparable magnetospheric compression that will open up to 60% of the Earth's geomagnetic field.

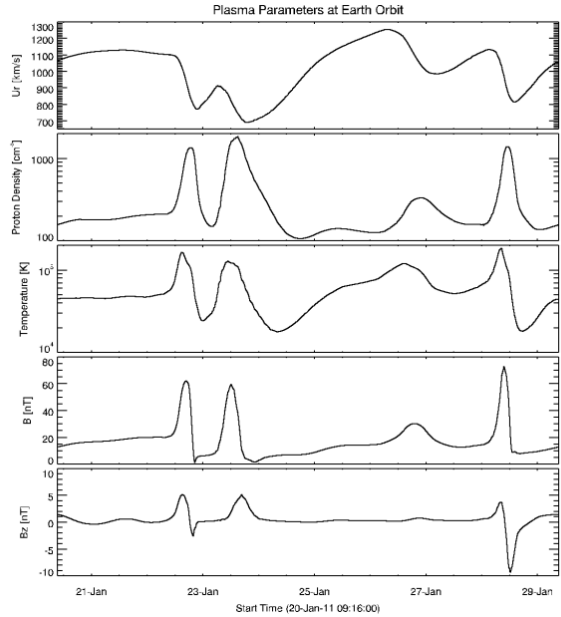

*Figure 8. The physical properties of the stellar CIR (2012.8 epoch) at Earth orbit: top to bottom panels represent the radial velocity (in km/s), proton density (in cm$^{-3}$), temperature (in K), magnetic field (in nT).*

*Figure 8* shows the enhancement of the plasma density in CIR driven shock up to 2000 cm$^{-3}$, which is by a factor of 100 greater than the CIR's density from the current Sun. The total magnetic field reaches 80 nT with the $B_z$ component of magnetic field of -10 nT, which is comparable to the values measured in solar CMEs introducing magnetic storms in the Earth's geospace (from magnetosphere to the middle atmosphere (Yurchyshyn et al. 2005; Li et al. 2018). The southward magnetic field of an CIR undergoes reconnection with the exoplanetary magnetic field and further pushes the magnetosphere close to the planet and opening the polar cap to a larger degree. We will study these geomagmetic field scenarios of interactions of stellar CIR driven shocks with Earth-like and giant exoplanets in our future studies. It is important to note that while these dense shocked regions can be important contributors to the transient magnetospheric compression of exoplanets, their contribution to the overall mass loss rate is small (a few percent of the stellar wind mass loss) because of their compact cross section areas.

Stellar CIR driven shock will also drive acceleration of energetic electrons and protons similar to solar CIRs (Richardson 2018) and would represent a quasi-steady structures swipping planetary magnetospheres of early Venus, Earth and exoplanets every 4.6 days lasting for ~0.5 days. Such shocks can form a population of accelerated particles up to 10 MeV that can serve as a As these particles penetrate into an $N_2$ - $CO_2$ rich atmospheres with a trace amounts of methane and water vapor, they can ignite the formation of complex molecules in mildly reducing atmospheres of early Earth and Earth-like exoplanets including hydrogen cyanide and formaldehyde, the precursors of proteins, complex sugars and building blocks of nulecobases (Patel et al. 2015; Airapetian et al. 2016; 2020; Rimmer & Rugheimer 2019).

4. Conclusions.

Here, we expanded the *AWSoM* stellar coronal model used previously to model stellar coronal environments by Alvarado-Gomez et al. (2016a,b) and Boro Saikia et al. (2020). These models utilized ZDI magnetograms of active stars, but did not use the constaned inputs for the thermodynamics of the stellar corona. Here, we used the fully thermodynamic MHD equations with the coronal heating source via input Alfvén wave energy flux specified at the upper chromosphere and contrained by HST/STIS observations of *κ$^1$ Ceti*. Thus, for the first time we have simulated the stellar corona as the full solution of the MHD chromospheric model. We also calculated the solar corona at CR 2106 to check that the model produces coronal and wind mass fluxes consistently with observations and for comparison with stellar models.

Our models find that the dominant dipole magnetic field derived from ZDI magnetic magnetograms of the star and the dissipation of the Alfven wave energy flux specified by the HST/STIS data produce a hot, 6 MK, corona, whichn is consistent with the observationally derived EM distribution from EUVE data (Teleschi et al. 2005). However, the EUV fluxes apeear to be 5-10 times smaller than observed fluxes. We suggested that this discrepancy can be attributed to the missing magnetic flux in small-scale magnetic structures associated with starspots to be added in the upcoming study. We also find that the global magnetic field of the star undergoes a transition to a tilted weaker dipole with developed multipole components 11 months later. This disorganized magnetic field produces the stellar corona at 6.3 MK, but with 40% less EUV coronal fluxes. The simulated mass loss rate from *κ$^1$ Ceti* at 2012.8 epoch appears to be 100 times greater than the mass loss rate observed from the current Sun's wind. We find that the weaker dipoled field at 2013.7 epoch produces the reduction of the mass loss rate by 40% .

A new structure simulated in our model is the stellar CIRs resulted from the interaction of fast and slow winds that has the dynamic pressure over 1380 times greater than that formed in the CIRs from the current Sun. This is a very important result as CIRs cause magnetic storms on Earth (Maggiolo et al. 2016; Richardson 2018). The initiation of magnetic storms is associated with enhanced dynamic pressure that compress the Earth's magnetosphere and cause the induction of ionospheric currents and results in Joule heating of the ionospheric and thermospheric layers of the planet. The enhanced magnetic field with the Bz component reaching 80 nT in our simulations is comparable to the magnetic fields of strongest CMEs. Such strong southward field can ignite magnetic reconnection with the (exo)planetary magnetospheric field, and thus produce strong fluxes of precipitating electrons and protons into the upper atmosphere provoking additional heating and associated enhanced (with respect to the EUV driven escape) escape rates. These processes can be crucial in evaluating the magnetospheric states of exoplanets around active stars because induced current dissipation will enhance the atmospheric escape from Earth-like exoplanets around active stars and can be critical for habitability conditions for rocky exoplanets in close-in habitable zones around red dwarfs (Cohen et al. 2014; Dong et al. 2018; Airapetian et al. 2020). We should note that these processes are complementary to more transient coronal mass ejection events that can increase the magnetospheric also contribute to larger atmospheric loss via their large dynamic pressures (Lugaz et al. 2015; Airapetian 2020).

Thus, our MHD model provides the framework to realistically model coronal environments of solar-like planet hosts by applying the data constrained setup developed for $\kappa^1$ *Ceti* to other young stars, specifically EK Dra, the "infant" proxy of our Sun at 100 Myr, younger solar-like (T Tau stars) at < 10 Myr and active K and M dwarfs. In conclusion, we should mention that the presented model accounts for the coronal heating in the closed magnetic loops only via Alfvén wave energy dissipation. However, recent high resolution observations of solar coronal loops suggest that some portion of the heating can be explained via explosive reconnection events known as nanoflares. In the future developments of this tool, we will introduce a heating term via nanoflares that contribute to the coronal loop and coronal background heating (Chitta et al. 2018; Bahauddin et al. 2020).

Our data driven MHD models can provide predictive capabilities for SXR and EUV intensities and the wind mass fluxes from young G, K and M planet hosts with constrained inputs supplied by TESS, HST, XMM Newton, NICER and ground-based facilities. The rotational modulation of the optical flux with TESS and other facilities will provide the starspot(s) size and its (their) locations that will be added to the ZDI magnetogram to account for the observed total unsigned magnetic flux. These reconstructions will be used to compare against observationally derived EUV fluxes (stars observed by EUVE observatory) and/or using their Far UV proxies (France et al. 2018). Thus, a coordinated multi-observatory campaign to address the stellar space weather environments is required to assess the habitability conditions of Earth-like planets around young Sun-like (G and K type) and M dwarfs. The first phase of the observational campaign "Evolving Magnetic Lives of Young Suns" has been recently started to characterize coronae, winds and superflares from young solar-like planet hosting stars (Soderblom et al. 2019).

Acknowledgements. V. S. Airapetian was supported by NASA Exobiology grant #80NSSC17K0463, TESS Cycle 1, HST Cycle 27 and NICER Cycle 2 grants. M. Jin acknowledges support from NASA's SDO/AIA (NNG04EA00C) contract to the LMSAL.O. Kochukhov acknowledges support by the Swedish Research Council, the Swedish National Space Agency and the Royal Swedish Academy of Sciences. The observations that support the findings of this study were obtained a a part of spectropolarimetric observations, previously obtained HST/STIS, MOST and TESS as well as XMM-Newton observations of $\kappa^1$ Ceti. The data are available in the public archive at https://mast.stsci.edu/portal/Mashup/Clients/Mast/Portal.html). This modeling work was performing by applying the SWMF/BATSRUS tools developed at The University of Michigan Center for Space Environment Modeling (CSEM) and made available through the NASA Community Coordinated Modeling Center (CCMC). Resources supporting this work were provided by the NASA High-End Computing (HEC) Program through the NASA Advanced Supercomputing (NAS) Division at Ames Research Center. Simulations were performed on NASA's HECC Pleiades cluster.